\documentstyle[preprint,aps]{revtex}
\begin{document}
\title {\bf COLOR  TRANSPARENCY AND THE VANISHING DEUTERIUM SHADOW}
\author{L.~L.~Frankfurt$^{a,e}$, W.~R.~Greenberg$^{b}$ , G.~A.~Miller$^{c}$,
 M.~M.~Sargsyan$^{a,f}$ , M.~I.~Strikman$^{d,e}$ }
\address{
$(a)$ School of Physics and Astronomy, Tel Aviv University, 69978, Israel, \\
$(b)$ Department of Physics,
University of Pennsylvania, Philadelphia, PA 19104 \\
$(c)$ Department of Physics, University of Washington, Seattle, WA 98195  \\
$(d)$ Department of Physics,
Pennsylvania State University, University Park, PA 16802\\
$(e)$ Institute for Nuclear Physics, St. Petersburg, Russia \\
$(f)$ Yerevan Physics Institute, 375036, Armenia }
\maketitle

\begin{abstract}
We find that the final state interactions in the $d(e,e'np)$ amplitude depend
strongly on the final momentum of the spectator nucleon. This means that color
transparency effects can be studied at rather low four-momentum transfer
Q$^2\ge~4~({\rm GeV/c})^2$ using un-polarized and polarized  deuteron targets.
\end{abstract}
\newpage
The Continuous Electron Beam Accelerator Facility CEBAF and
HERMES (DESY)
are about to begin running experiments. These
new installations offer unprecedented
luminosity and a
continuous beam\cite{cebafdoc,hermesdoc}. Several electron-deuteron
scattering  experiments
(including some with polarized targets) are planned to study
$ed$ reactions
at  CEBAF and HERMES. The new features of these experiments is the
ability to detect also the final state nucleons (hadrons) in coincidence with
scattered electrons covering transfered energy range
1 (GeV/c)$^2\leq$ Q$^2\leq~20~({\rm GeV/c})^2$.

We point out that these very same experiments can be used to gain insight into
how quantum chromodynamics  influences nuclear interactions at fairly low
values of Q$^2$. To understand this, recall some well known properties of the
deuteron. First,
the deuteron is the best understood nuclear system, with a wave
function determined experimentally in a wide momentum range \cite{Brown}.
Second, it has long
been known that the total cross section for projectile-deuterium
scattering deviates from
the sum of the neutron and proton cross sections because one
of its particles
sometimes lies in the shadow cast by the other\cite{Glauber}. The
earliest estimates are that
\begin{equation}
\sigma_d=\sigma_n +\sigma_p -{\sigma_n\sigma_p \over 4\pi}
<d|{1\over r^2}|d>,
\label{sh}
\end{equation}
where $r$ is the
operator representing the internucleon separation and $|d>$ is the
deuteron wave function. The
${1\over r^2}$ behavior causes the second scattering to
occur for small
distances $r$ despite the typically large ($\approx$ 4 fm) separation
between the $n$ and the $p$.  This is exploited below.

In QCD, the absorption
of a high momentum virtual photon by a nucleon leads to the
production of a small-sized color
singlet state for sufficiently high values of Q$^2$.
We optimistically term the
small-sized wave packet a  point like configuration PLC.
Such a state would not interact
with the spectator nucleon, and the deuteron would
lose its shadow.  This
vanishing of a final state interaction (FSI) has been termed
color transparency (CT).

Despite intense experimental and theoretical investigation no unambiguous
evidence for this novel effect has been observed. It is our view that a PLC is
actually produced, but that it  expands as it propagates through the
nucleus\cite{ANN}. The expanded system interacts strongly and obscures the
physics of the initial PLC. The time  or distance required for the expansion is
of the order of $l_h\sim 0.4 (p$/GeV) fm, where $p$ is  the momentum of the
PLC. If $l_h$ is greater than the nuclear radius the expansion effects are
minimal.
This condition, which can be achieved for very high values of the momentum
transfer, has
not yet been met in an experiment.

The precise d(e,e'pn) experiments
we discuss could provide the long-sought signature
of CT. The deuteron wave function peaks for $r\approx 1.8$ fm, and the
relevant distances for rescattering are determined
by  operators which include a
${1\over r^2}$ dependence. Thus for
the deuteron (and other light nuclei) the PLC needs
only to remain of small size for short propagation distances.

Suppose an
incident virtual photon of four momentum ($\nu, \vec q$) leads to the
detection of an
outgoing nucleon with the large momentum $\vec p_f=\vec q -\vec p_s$,
and the other ``spectator"
nucleon of momentum $\vec p_s(p_s^z,p^t)$, in which the
subscript $z$ and $t$
denotes a direction parallel and perpendicular to that of
$\vec q$.  The
importance of FSI is maximized by using
so-called ``perpendicular
kinematics"\cite{Eli},
in which the light cone fraction of the deuteron momentum carried by a
spectator nucleon
(mass m) $\alpha\equiv {\sqrt{m^2+p_s^2}-p_s^z\over m}\approx~1$, but $p_t$ is
not negligible.
Then the  spectator
momentum $\vec p_s$ is approximately perpendicular to  $\vec q$.

The scattering amplitude
${\cal M}$, including the $np$ final state interaction, can be written
using the eikonal apporximation as
\begin{eqnarray}
{\cal M} = <p_s^z,\vec p_t|d> -{1\over 4i}\int{d^2k_t\over
(2\pi)^2}<\tilde p_s^z,\vec p_t-\vec k_t|d>\times\nonumber \\
 f^{np}(\vec k_t)\left[1-i\beta\right],
\label{eq:amp}
\end{eqnarray}
where
$\tilde p_s^z=p_s^z-(E_s-m){M_d+\nu\over |\vec q|}$, $E_s=\sqrt{p_s^2+m^2}$
and $M_d$ is the mass of the deuteron. Spin indices and factors arising from
the
electron-nucleon interaction are suppressed to simplify the notation.
The factor $\beta$ accounts for theta function of the eikonal
Green's function.
However when  $\alpha\rightarrow 1$ $\beta\approx 0$\cite{FGMSS94}.
The function $f^{np}$ represents the FSI between the outgoing
nucleons. We use a parametrization
 \begin{equation} f^{pn} =
\sigma^{pn}_{tot}(i+a_n)e^{-b_n k_t^2/2}, \label{eq:fnp} \end{equation} for the
$np$ scattering amplitude. The quantities $\sigma^{pn}_{tot}$, $a_n$ and
$b_n$, at Q$^2>3$ (GeV/c)$^2$ depend weakly on the momentum of the knocked-out
nucleon with $\sigma^{pn}_{tot}\approx 40$ mb, $a_n\approx -0.2$ and
$b_n\approx~6-8$ GeV$^{-2}$ for the kinematics of our interest.

The sensitivity to CT effects that we observe rests on the very
different $\vec p_t$ dependence of the two terms of Eq.~(\ref{eq:amp}) at
$\alpha\approx 1$. What can one expect? For $\vec p_t\approx 0$, the ratio of
the second term to the first is of the order of $-\sigma^{pn}_{tot}/16\pi
R_d^2$ and is small and negative. Thus, at low $p_t$, final state interactions
reduce the value of the computed cross section. This is the shadowing effect
mentioned above. But the first term falls more rapidly than the second as the
magnitude of $\vec p_t$ increases. This is because the fall off is controlled
by the large deuteron size in the first term and by the small range of the $np$
interaction in the second term. (In the limit of zero range ($b_n\to 0$), the
second term is proportional to $\int d^3 r {1\over r^2}<\vec r|d>$.)  As $p_t$
increases  from zero, the relative importance of the shadowing grows. However,
if $p_t$ is further increased, the value of $|{{\cal M}\over <\vec p_t|d>}|^2$
actually increases!

We define the transparency $T$ as the ratio of the measured cross section
(or calculated cross section with FSI) to the one calculated in the plane wave
impulse approximation (PWIA):
\begin{equation}
T(Q^2,\vec p_t,\alpha) \equiv {\sigma^{FSI}_{d(e,e'pn)}
(Q^2,\vec p_t,\alpha) \over \sigma^{PWIA}_{d(e,e'pn)}(Q^2,\vec p_t,\alpha).}
\label{T1}
\end{equation}
Fig.~1 shows the  dramatic dependence of $T$ on the magnitude of $\vec p_t$ as
a function of Q$^2$ for an unpolarized target.

At $p_t~\leq~200$ MeV/c the FSI lead to shadowing,
with $T(p_t=0)\approx 0.97$ and
a much smaller
$T(p^t_{n}=0.2$ GeV/c) $\approx  0.5$.
But for $p_t~>~300$ MeV/c one finds $T>1$.
These features are
apparent for all Q$^2\geq~2$ (GeV/c)$^2$, but are more significant
for the
larger values of Q$^2$. Other kinematics are examined in Ref.\cite{FGMSS94}.

Including the effects of CT would change the results of Fig.~1.
For
sufficiently large
Q$^2$ the final state interactions would be eliminated entirely.
But for values of Q$^2\leq~10$ GeV$^2$, the deuteron is not completely
transparent. Our calculations must include
the effects of PLC expansion. We use two
models which
account for the formation of the PLC and their evolution to the normal
hadronic state:
the quantum diffusion model \cite{FLFS} and the three state model of
Ref.\cite{FGMS93}. For both models, the parameters we use
are in the range consistent with the (p,2p)\cite{carroll}
and SLAC (e,e'p)\cite{ne18} data.

The reduced interaction  between
the PLC and the spectator nucleon can be described
in terms
of its transverse size and distance $z$ from the photon absorption point.
In the quantum diffusion model the PLC-N scattering amplitude takes the
form\cite{EFGMSS94}:
\newpage

\begin{eqnarray}
f^{PLC,N}(z,k_t,Q^2) & = & i\sigma_{tot}(z,Q^{2}) \cdot
e^{{b_n\over 2 }t}\times\nonumber \\
 & & {G_{N}(t\cdot\sigma_{tot}(z,Q^{2})/\sigma_{tot})
\over G_{N}(t)},
\label{F_NNCT}
\end{eqnarray}
where t$= -k_t^2 $, and $G_{N}(t)$ is the Sachs form factor.
In Eq.~(\ref{F_NNCT}) $\sigma_{tot}(z,Q^{2})$  is the  effective
total cross section of the  interaction  of the PLC at the distance $z$
from the interaction point. This is~\cite{FLFS}:
\begin{eqnarray}
\sigma _{tot}(z,Q^{2})  & = & \sigma_{tot}^{pn}
\left \{ \left ({z \over l_{h}} +
{\langle r_{t}(Q^2)^{2} \rangle \over \langle r_t^{2}  \rangle }
(1-{z \over l_{h}}) \right )\Theta (l_{h}-z)\right.\nonumber \\
 & &  +  \left.\Theta (z-l_{h})\right\},
\label{SIGMA_CT}
\end{eqnarray}
where $l_h = 2p_n/\Delta~M^2$, with $\Delta~M^2=0.7$ GeV$^2$.
Here ${\langle r_{t}(Q^2)^{2} \rangle}$  represents the transverse size of the
initially produced configuration. Several realistic models indicate\cite{FMS92}
that this is negligibly small for  Q$^2~\geq~1.5$ GeV$^2$.

The three state
models, which allows also the computation of resonance production
cross sections, is
based on the assumption that the hard scattering operator acts
on a nucleon
to produce a non-interacting $|PLC\rangle$ which is a superposition
of three baryonic states:
\begin{equation}
|PLC\rangle = \sum_{m=N,N^*,N^{**}} F_{m,N}(Q^2) | m \rangle,
\label{TH}
\end{equation}
where $F_{m,N}(Q^2)$  are elastic ($m=N$) and inelastic transition form
factors. We assume that all form factors have the same Q$^2-$dependence and
also neglect possible spin effects in the form factors. CT is introduced by
the statement that the initially produced PLC undergoes no FSI\cite{FGMS93},
\begin{equation}
T_S|PLC\rangle = 0,
\label{SUM}
\end{equation}
where $T_S$ is the matrix representing the soft final state interactions.
$T_S$ is represented by the most general 3 $\times$ 3 Hermitian matrix
consistent with Eq.~\ref{SUM}. We use $T_S$ of Ref.~\cite{FGMS93}, with
the parameters $M_N^*$=1.4 GeV, $M_{N^{**}}$=1.8 GeV, $\epsilon$ =0.17,
$F_{N,N}/F_{N,N^{**}}$=1.0, $F_{N^*,N}/F_{N,N^{**}}$=3.1.

We compare the predictions of these two models of CT in Fig.~2.
The ratios of quantities T of Eq.~(\ref{T1}) computed with $FSI$ according
to CT - T$^{CT}$ or according to the usual Glauber approximation
- T$^{GA}$ are shown. We find T$^{CT}$/T$^{GA}>1$ for $p_t\leq 200$ MeV/c,
and $T^{CT}/T^{GA}<1$ for $p_t~> 300$ MeV/c.

One may also compute and measure ratios of cross sections for different values
of $p_t$. This quantity represents the ratio of directly measured  experimental
quantities, and does not require   additional normalization to the
corresponding PWIA calculation (as in Eq~(\ref{T1})).  A study of Fig.~2 shows
that the effects of color transparency can modify such ratios by as much as
$30\%$ for Q$^2$ as low as 6-10 (GeV/c)$^2$.

We next
discuss the possibility of  using a polarized deuteron target to investigate
color coherent
effects. Using different target polarizations emphasizes the role of
the deuteron $d$-state causing smaller space-time intervals to be probed.
For numerical estimates we consider the  asymmetry $A_d$ measurable in
electrodisintegration of the polarized deuteron:
\begin{equation}
A_d(Q^2,\vec p_s) = {\sigma(1)+\sigma(-1)-2\sigma(0)
\over \sigma(1)+\sigma(0)+\sigma(-1)}
\label{asm}
\end{equation}
where  $\sigma(s_z)~\equiv~{d\sigma^{\vec s,s_z}\over dE_{e'} d\Omega_{e'}
d^3p}$, $s$ and $s_z$ are the spin and it's  $z$ component of the deuteron.

It is useful to recall some properties of the  s- $u(k)$ and d- $w(k)$
state wave functions in momentum space. The quantity $u(k)$ decreases as $k$
increased from 0, and changes sign at $k\approx~400$ MeV/c, while $w(k)$ grows
with $k$ from a negative minimum at $k\approx~100$ Mev/c. Thus in  some range
of momenta, the $w(k)$  is comparable to  (or larger) than $u(k)$ (for
details  see \cite{Brown}).
These well-established features cause
the tensor polarization (the numerator of  Eq.~(\ref{asm}))
to be comparable to the unpolarized cross section (denominator of
 Eq.~(\ref{asm})).
In particular,
at $p_t\approx 300$ MeV/c and in the PWIA the asymmetry calculated
according to Eq.~(\ref{asm}) is close to unity\cite{FGMSS94}.
Deviations from unity originate predominantly from the effects of FSI.

We present the Q$^2$ dependence of the asymmetry $A_d$  for ``perpendicular"
kinematics, at $p_t=300$ MeV/c. This figure clearly demonstrates the computed
importance of CT effects.

The reliability of
our interpretation of an experimental measurement depends on the
dominance of FSI  in causing deviations of $T$ and $A_d$ away
from the
plane wave results. We claim that competing effects are  restricted to small
values by the chosen
conditions:
(a) only small nucleon momenta in the deuteron
$\leq~300-350$ MeV/c are  relevant here,
(b) perpendicular kinematics, $\alpha\approx 1$,
(c) the observables are the ratios of
experimental quantities in (nearly) similar
kinematical conditions,
(d) the FSI amplitude of Eq.~(\ref{eq:amp}) is dominated by small
values of nucleon Fermi momenta.

The conditions (a) and (b) are
sufficient to suppress relativistic effects of nucleon
motion in the initial state.
One measure of such effects is the difference between the
nucleon Fermi momenta defined
in the light cone\cite{FS81} and nonrelativistic theories
of deuteron: $\sqrt{{m^2+p_t^2\over \alpha(2-\alpha)}-m^2}-\sqrt{k_z^2 + p_t^2}
\mid_{\alpha\rightarrow 1} \approx {p_t^3/8m^2}$. This is small if (a) and (b)
are satisfied.
The deuteron wave functions are well-known if condition (a) holds\cite{Brown}.
Relativistic effects and nucleon binding effects were examined closely in
Ref.\cite{FGMSS94} and   the overall effects are no more than
a few percent of
$T$  and $A_d$.

The influence
of meson exchange currents (MEC) and $\Delta$-isobar contributions
(IC)\cite{Aren}  are mechanisms
which are potentially competitive with the influence
of FSI. But, for our kinematics MEC are suppressed by the structure
of  the $\gamma^*N\rightarrow N\pi$ transition
matrix. Such transitions are related to
the nucleon sea quarks. These are not very important for Q$^2\geq 1$ GeV$^2$
and
$x_{Bj}\sim\alpha\sim 1$ where valence quarks dominate.

The role
of IC is also expected to be small, because the $\gamma^*N\rightarrow\Delta$
transition
form factor decreases more rapidly with Q$^2$ than elastic nucleon form
factors\cite{STL}. The
$\Delta$ contributions are further suppressed because the
$\Delta N \rightarrow NN$ amplitude
is predominantly real and decreases rapidly with
energy (since it is dominated by
pion exchange) whereas the FSI effects we study are
determined by the
imaginary part of the soft rescattering amplitude. Another FSI channel
involves
the final state pion charge-exchange (CHE) interaction. In our kinematics
$-t\geq 0.05$ (GeV/c)$^2$. The charge-exchange amplitude
drops more strongly with $-t$ than the elastic amplitude, causing the
correction to be small
(see e.g. \cite{ERIC}).

To estimate MEC, IC, and CHE contributions
numerically the predictions of Glauber
approximations were
compared in Ref.\cite{FGMSS94} with the predictions of the model
of Arenh\"{o}vel et al.\cite{Aren2}, which includes all of those effects. For
perpendicular
kinematics, the combined influence of MEC, IC and CHE  cause only
$\sim 10\%$ changes
in computed values of $T$ for Q$^2=1$ (GeV/c)$^2$ and  $p_t=400$ MeV/c.
These contributions even smaller for larger values of Q$^2$ and $p_t$.

For perpendicular kinematics,
the combined influence of the various
correction terms discussed above
are typically no more than $\sim 5\%$ and $\sim 10\%$, at
Q$^2>2$ (GeV/c)$^2$, for unpolarized $T$ and  polarized measurements $A_d$.
This is significantly smaller than the effects of CT that we predict.

The use of ``perpendicular" kinematics significantly increases the sensitivity
to FSI and allows smaller  than average internucleon distances to be probed.
Depending on whether the Born or rescattering term dominates the Glauber
approximation to the scattering amplitude, CT effects cause either an increase
or decrease of the $d(e,e'pn)$ cross section. Based  on this, we suggest that
the ratios of $d(e,e'np)$ cross sections at $\alpha\approx 1$, $p_t\sim 200$
MeV/c
and at $\alpha\approx 1$, $p_t\sim 300$ MeV/c be measured. Our calculations
of this
ratio predict 20-40\% effects for Q$^2\sim 4-10$ (GeV/c)$^2$. We also suggest
that
the tensor asymmetry in the $\vec d(e,e'pn)$ reaction be measured. Here the
conventional Glauber approximation predicts a huge FSI and the CT models
predict a
50 - 100\% change in the asymmetry at Q$^2\sim 4-10$ (GeV/c)$^2$.
The measurements
we suggest could provide definitive evidence for or against color transparency.

This work is supported in part by the U.S. Department of Energy and BSF.



\begin{figure}
\caption{$T$ of Eqs.~(4) and (2), $\alpha$ = 1, $\vec p_s \approx \vec p_t$.}
\end{figure}

\begin{figure}
\caption{$p_t$ and $Q^2$ dependence of ratios of $T^{GA}/T^{CT}$ of
Eq.~(4), $\alpha$ = 1. a) quantum diffusion, b) three state model.}
\end{figure}

\begin{figure}
\caption{$A_d(Q^2,\vec p_s,\vec p_t)$ of Eq.~(9).  Solid line - GA,
dashed line quantum diffusion model, dash-dotted - three state model,
dotted line PWIA. $\alpha$ = 1, $p_t$ = 300 MeV/c.}
\end{figure}

\end{document}